\documentclass[prl, amsmath, amssymb, preprint]{revtex4-1}

\usepackage{graphicx}
\usepackage{dcolumn}
\usepackage{bm}

\usepackage{hyperref}
\usepackage{subcaption} 
\usepackage{amsmath} 
\usepackage{float}
\usepackage[section]{placeins}

\begin{document}
\title{Optical Trapping of Large Metallic Particles in Air}

\preprint{prl/123-QED}

 \author{S. Mirzaei-Ghormish}
 \affiliation{Department of Electrical and Computer Engineering, Brigham Young University, Provo, UT}
 \author{S. Griffith}%
 \affiliation{Department of Electrical and Computer Engineering, Brigham Young University, Provo, UT}

 \author{D. Smalley}
\affiliation{Department of Electrical and Computer Engineering, Brigham Young University, Provo, UT}
 \email{smalley@byu.edu.}

 \author{Ryan M. Camacho}
\affiliation{Department of Electrical and Computer Engineering, Brigham Young University, Provo, UT}

\date{\today}

\begin{abstract}

In this paper, we introduce a horizontally-oriented, photophoretic  `boat' trap that is capable of capturing and self-loading large (radius ${\ge}1$ $\mu$m) solid gold particles in air for more than one hour.  Once trapped, particles are held stably, even as the trap is modified to scan axially or expanded to a larger size to increase the capture cross-section. We theoretically present and experimentally demonstrate each of these affordances.  We describe the utility of such to investigate large, metallic, and plasmonic particles for display applications. 
\end{abstract}

\maketitle

\section{Introduction}
Photophoretic displays have begun to meet the challenge presented by the century-long search for free-space volumetric images \cite{smalley2018photophoretic}.  However, to make these displays practical, the trapped particles would ideally have strong and broadband light-matter interactions along with a large capture cross-section and singular trap locus.  Trapping large metallic particles may be ideal for this and other applications; however, it has proven notoriously difficult to robustly trap large metallic particles in air. 

For large metallic particles exposed to laser light, optical forces are mainly driven by positive photophoretic forces. Unlike smaller metallic nanoparticles, the plasmonic absorption spectrum shifts to longer wavelengths and broadens \cite{brzobohaty2015three,KHLEBTSOV20101}, causing uneven heating on the illuminated particle side. Due to the Soret effect, this creates a positive photophoretic force that pushes the particle from hotter to colder regions, overpowering other forces. Consequently, these particles are trapped at intensity minima, unlike smaller or less absorbent particles that can be trapped at intensity maxima due to negative photophoretic or gradient forces \cite{JOVANOVIC2009889}.
 
All known efforts to optically trap metallic particles away from surfaces are shown in Fig. \ref{boat}(a)\cite{svoboda1994optical,seol2006gold,hansen2005expanding,brzobohaty2015three,jauffred2015optical,arita2018invited,ashkin1974stability,roosen1978tem,sasaki1992optical,sato1994optical,furukawa1998optical,ke1999characterization,t2000three,gu2002three}. We calculate the photophoretic force and extinction cross-section for each particle based on its reported size and material. The extinction cross-section encompasses both scattering and absorption, contributing to repulsion. Data point size reflects the particle size, red markers indicate air-trapped particles, and blue markers indicate water-trapped ones. Calculations assume spherical shapes, though non-spherical shapes are expected to reduce repulsive forces\cite{brzobohaty2015three}.

Several key observations can be made: First, larger metallic particles experience greater repulsive radiation and photophoretic forces as anticipated. Second, photophoretic forces decrease when particles are in water. Notably, all trapped metallic particles do date have operated in regimes where attractive forces prevail or thermal effects are negated, highlighting the need for new air-trapping geometries for large metallic particles.

 Further details reveal that metallic particles have been trapped in both water and air using inverted optical tweezers in the dipole regime (radius $\ll$ wavelength)\cite{svoboda1994optical,seol2006gold,hansen2005expanding,brzobohaty2015three,jauffred2015optical,arita2018invited}. Here, trapping relies on attractive gradient forces, overshadowing repulsive scattering and thermal forces. However, such methods can dramatically heat the particle surface beyond the melting point, limiting utility. In the Mie regime (radius $\geq$ wavelength), repulsive forces may destabilize the particles. Reflective coatings can mitigate Soret effects \cite{ashkin1974stability,roosen1978tem}. Mie particles have also been trapped in 2D or loosely in 3D with engineered beams in water, with size limitations \cite{sasaki1992optical,sato1994optical,furukawa1998optical,ke1999characterization,t2000three,gu2002three}. Trapping near extreme fields at surfaces has also been reported \cite{zhang2021plasmonic}. For display applications requiring free-space axial and radial movement, new approaches to air trapping of large metallic particles are needed.

We finally note that several previous experiments have successfully trapped non-metallic absorbing particles photphoretically using optical beams with local intensity aberrations \cite{Shvedov:10, Shvedov:11, Zhang:12, gong2016}, leading to demonstrations of 3D volumetric displays \cite{smalley2018photophoretic}. However, photophoretic forces in similarly sized metallic particles are 1-2 orders of magnitude larger, requiring more robust trapping methods. Moreover, the density of metallic particles is much larger than that of non-metallic particles (carbon, soot, black liquor). Therefore, the local trap sites created by speckle and aberration are unable to hold the metallic particles.  We attempted to trap both soot and gold particles using these methods. While the soot particles were stably trapped, we didn't observe a stable trapping for gold particles. 

In this work, we trap large gold particles in air for more than one hour, demonstrating for the first time the trapping of large metallic particles in air without proximity to a surface.  Our trap design builds on previous work demonstrating arbitrary acousto optic control of optical trapping beams \cite{friedman2000compression,milner2001optical} and a resulting 3D morphology that allows for near-solid state axial and lateral translation \cite{gong2018optical}.  Our design also has an unexpected and convenient photohoretic (rather than gravitational) self-loading phenomenon not previously reported, and the ability to dynamically change the trapping potential and location [Fig. \ref{boat}(a)].  This enables, simultaneously, a large capture cross-section and a singular stable trap locus with an overall action reminiscent of loading within a mechanical pencil. The dynamic nature of the trap also allows for adjustment of the absolute location of the self-loading terminus--providing a possible solution to the challenge of variability in other self-loading systems \cite{kirian2016development}. To explain the underlying physics of our trap, we present a theoretical model of the dynamical trap in terms of all relevant optical forces.  Using our model, we calculate the trapping potential in the transverse and propagation planes and show the self-loading operation.  Our theory is in good agreement with experimental observations. 

\section{Trap design}
\begin{figure}
    \centering
     \includegraphics[width=.48\textwidth]{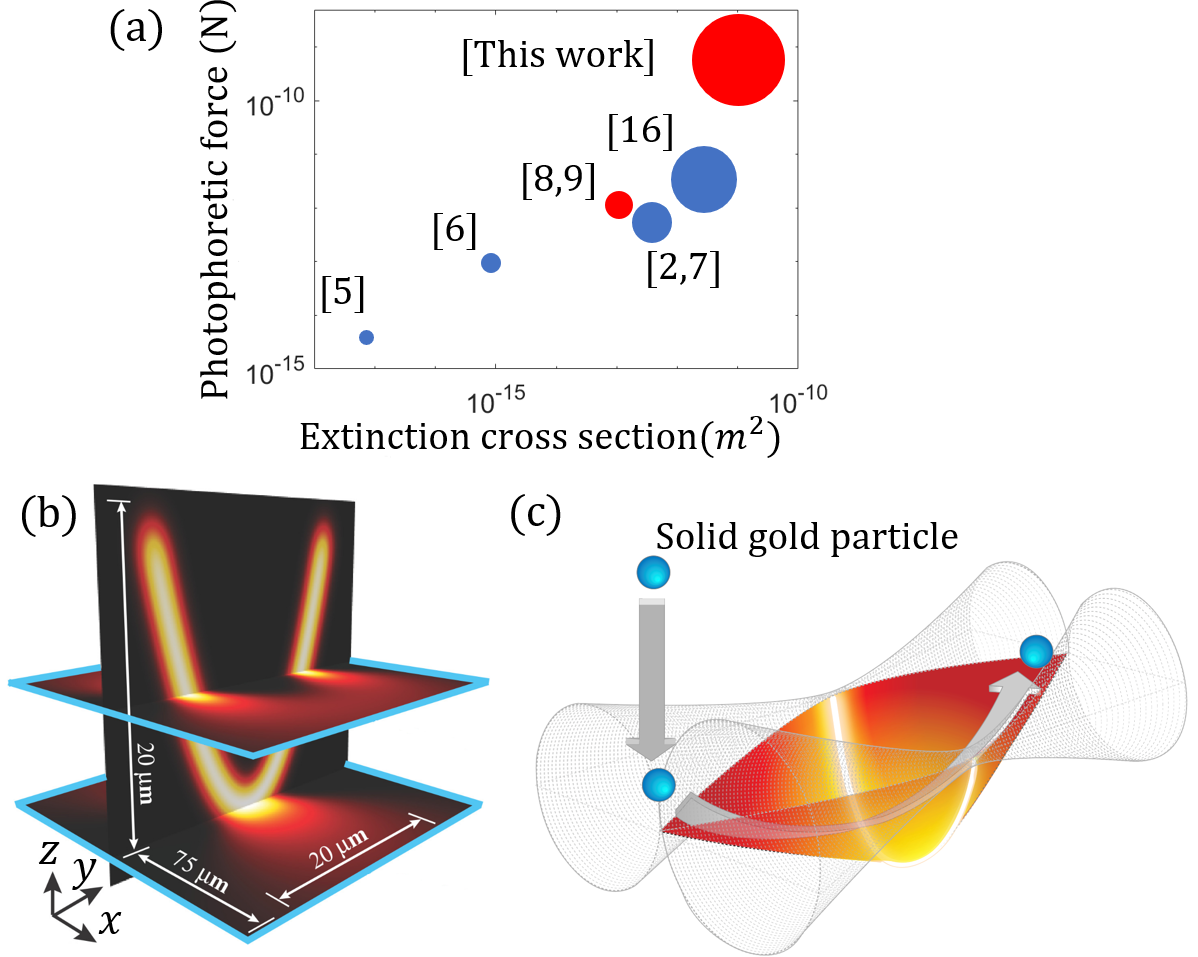}
     \caption{(a) Illustrated contribution of this work by comparing the calculated photophoretic force and extinction cross-section of published experimental optical trapping of metallic particles in air and water. (b) The three-dimensional boat-shaped intensity distribution of the trap reported in this work. (c) This boat trap shows a self-loading operation for metallic particles. }
    \label{boat}
\end{figure}

To achieve a 3D photophoretic trap for metallic particles, we form a dynamic repulsive energy well by rapidly scanning a focused laser along a parabolic path in the $y$-$z$ plane, as depicted in Fig. \ref{boat}(b). If the scan rate exceeds the particle's diffusional motion, the particle senses a quasistatic, repulsive intensity distribution. Details on achieving uniform side walls are in the Supplemental Material \cite{supp}.

It is straightforward to see that the scanning laser forms a parabolic well in the transverse plane. What's less obvious but crucial is how it also creates a deep trapping potential along the $x$-axis. As illustrated in Fig. \ref{boat}(b), the quasistatic electric field has two nearly parallel diverging Gaussian beams whose intensity resembles a boat shape. The intensity peaks along the $x$-axis are due to the overlap of these beams at two points: the stern and the bow of the "boat." We call these convergence points, and they repel particles to the boat's center. Although the electric field is $x$-axis symmetric, the photophoretic force is not, as the laser illuminates the particle from one side. Interestingly, due to beam divergence, the opposite side of the particle can be preferentially illuminated at the positive convergence point. This enables a single laser to push a particle in the $-x$ direction, despite propagating in the $+x$ direction. This also implies that the trap's potential minimum will be near this positive convergence point. The superposition of all beams results in a 3D boat-shaped potential, shown in Fig. \ref{boat}(c).

The intensity in the transverse plane at time $t$ may be expressed as $z_f(t)=\left(\frac{2}{d}\right)y_f^2(t)-d$ where$\quad -d \le y_f(t) \le d$ is the tuning parameter that controls the size of the potential trap in both transverse and axial planes. 
The dynamic intensity is
\begin{equation}
    I(x,y,z;t) = \left(\frac{2p_{in}}{\pi w^2(x)}\right)e^{-2\left(\frac{(z-z_f(t))^2+(y-y_f(t))^2}{W^2(x)}\right)},
    \label{equation2}
\end{equation}

where $w(x)=w_0\sqrt{1+\left(\frac{x}{x_0}\right)^2}$ is the beam width,  $w_0=\frac{2\lambda}{\pi NA}$ is the beam waist, $NA$ is the numerical aperture, $x_0=\frac{k_0 w_0^2}{2}$ is the Rayleigh length, and $p_{in}$ is the laser power and the focal point of the laser beam is located at the $x=0$ plane.  To create a uniform intensity, the speed of the scanner is adjusted to be constant along the parabolic path as described in the Supplemental Material \cite{supp}.

Particles are trapped inside the boat's dark region by repulsion from the surrounding intensity barrier. The trap is adiabatic and its volume can be adjusted using the parameter $d$, controlled by the AOMs' radio frequency. The dimensions of the boat—width  $W=2d$, depth $H=2d$, and length $L=(k_0 \omega_0) \sqrt{{y_1^2}-{w_0^2}}$ (where $y1=\sqrt{\frac{d}{2}(w_0+R)}$)—are specified in Fig. S1 \cite{supp}. Both width and height depend solely on $d$, while length also depends on the numerical aperture. Increasing $d$ expands the boat's volume, and a higher $NA$ compresses it axially. A larger boat holds more particles, while a smaller one offers a deeper potential.

\section{Theoretical Calculation of Trapping Forces}

We model the boat trap by calculating all relevant optical and thermal forces and their corresponding energy potentials. The effect of Langevin forces including the Brownian motion is considered by comparing the depth of the potential traps with thermal energy $k_{B} T$. Here we neglect the Drag force, as it is small compared to the optical and thermal forces. The details of the calculations are shown in the Supplemental Material \cite{supp}.

  \begin{figure}
    \centering
     \includegraphics[width=.48\textwidth]{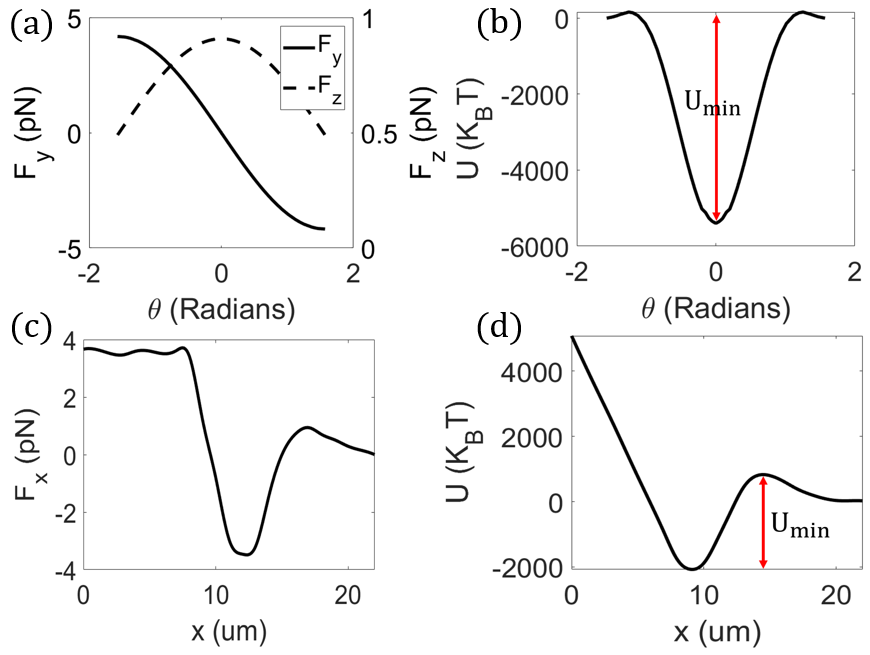}
     \caption{ (a) The $y$ component and the $z$ component of the total force, and (b) the potential trap in the transverse plane. (c) The total force and (d) the potential trap in the longitudinal direction. The simulation parameters are chosen as follows:  $\lambda=532nm$, $p_{in}=10$ mW, $NA=0.15$, $\omega=5 $kHz, and $d=5$ $\mu$m. The optical forces are found by numerically calculating the Maxwell stress tensor Eq. S4, and the photophoretic force is calculated using Eq. S7, both in the Supplemental Material \cite{supp}.}
    \label{force}
    
\end{figure} 

In the transverse plane, we calculate the total forces (optical and photophoretic forces) and potential trap on the line  $z=\left(\frac{2}{d}\right)y^2-d+(\omega_0+R)$. This path follows the $1/e^2$ intensity point of the Gaussian beams relative to the $y$-$z$ plane, offset by the radius of the particle,
In Fig. \ref{force} (a and b), the total force in the $y$ and $z$-directions and the transverse potential trap are shown as a function of angle $\theta$ at $x=0$. Since the $y$ component of the total force is an odd function of $\theta$, where it is negative for $0\le\theta\le\frac{\pi}{2}$ and positive for $-\frac{\pi}{2}\le\theta\le0$, the particle is continuously pulled into the trap region in the $y$-direction. In contrast, the $z$ component of the total force is an even function of $\theta$ and its values are positive.  This means that the particle experiences the repulsive forces in the $z$-direction that overcomes gravity. The vector map of the total force in the transverse plane is shown in the Supplemental Material \cite{supp}. The potential energy of the trap at point $\rho=(y,z)$ is  

\begin{equation}
    U(\rho)=-\int_{\infty}^{\rho}\vec{F}_{tot}\cdot d\vec{\rho}=-\int_{\infty}^{y}\left( F_y+\frac{4}{d}F_zy' \right)dy'
    \label{equation12}
\end{equation}

The potential \(U(\rho)\) is depicted in Fig. \ref{force}(b), showing a symmetric trap in the transverse plane with a minimum at \(\theta=0\) (\(y=0\)). To counteract Brownian motion, the trap depth \(U_{\text{min}}\) is \(5500k_B T\) (ten times thermal energy), ensuring stability. Similar calculations for successive x values yield a 3D potential \(U(x,y,z)\).

In the axial direction, we calculate the potential along a path in the $x$-$z$ plane at $z=w(x)+R$. This path aligns with the $1/e^2$ intensity point of the Gaussian beams and is offset by the particle's radius. Due to computational expense, we focus on electric field regions near the particle. Given that the beam size significantly exceeds the particle size ($w_{0}>>R$), we approximate the field distribution as the sum of three discrete beams, shown in Fig. S1 \cite{supp}. Their focal points are at $(0,y1, -d+w_0+R)$, $(0,0,-d)$, and $(0,-y1,-d+w_0+R)$, converging axially at $x= \pm13\mu m$.

In the axial direction, we evaluate the potential along a path at \(z=w(x)+R\), aligned with the \(1/e^2\) intensity of the Gaussian beams and offset by the particle radius. To reduce computational load, we focus on nearby electric field regions. The beam size vastly exceeds the particle size (\(w_0 >> R\)), allowing us to approximate the field as the sum of three discrete beams, detailed in Fig. S1 \cite{supp}. Their focal points are at \((0,y1, -d+w_0+R)\), \((0,0,-d)\), and \((0,-y1,-d+w_0+R)\), converging at \(x= \pm13\mu m\).

Then, the potential energy of the trap at point $(x,z)$ is
\begin{equation}
    U(x)=-\int_{\infty}^{x}\left( F_x+\frac{\omega_0^{2}}{x_0^{2}}\frac{x'}{\omega(x')}F_z \right)dx'
    \label{equation13}
\end{equation}

 Figure \ref{force} (c and d) illustrates the axial force and potential trap. For \(x<0\), the force is positive, indicating repulsion toward the focus. For \(x>0\), the force initially remains positive but turns negative near the positive convergence point. This negative force in the \(9\mu m \leq x \leq15 \mu m\) region leads to particle trapping. The resulting potential is asymmetric, centered near \(x=9 \mu m\), and has sufficient depth \((U_{min}=2500k_B T)\) to counter Brownian motion, ensuring stable trapping. Combined with transverse potentials shown in Figs. \ref{force}(b and d), the equilibrium lies near the positive convergence point near the 'center-bow' of the boat.

\section{Experimental demonstration}

\begin{figure*}
    \centering
     \includegraphics[width=\textwidth]{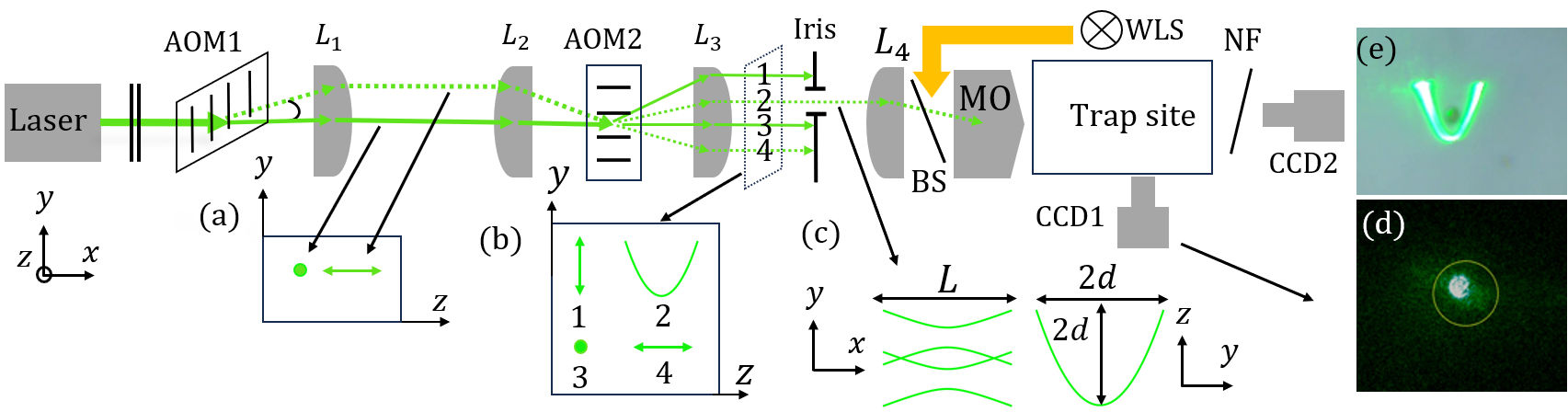}
     \caption{Schematic diagram of the experimental setup from top view. The dynamic boat-shaped intensity is created by a pair of AOMs, where the first AOM scans light in the z-direction and the second AOM in the y-direction.}
    \label{experimentalSetup}   
\end{figure*}

Experiments were performed to validate the theoretical predictions of the previous section and demonstrate the advantages of the proposed boat-shaped trap. A top view of the experimental setup is shown in Fig. \ref{experimentalSetup}.
A linearly polarized Gaussian beam derived from a CW coherent laser source (Verdi-V10, $\lambda=532nm$) passes through a pair acousto-optic modulators (AOMs). The AOMs are driven by a Rigol DG4202 arbitrary waveform generator with a voltage-controlled oscillator (VCO) and amplifiers. The incoming light passing through AOM1 is diffracted, where the zero-order (solid line) and first-order (dashed line) diffracted beams have the highest diffraction efficiency. The zero-order is an undeflected point beam, and the first-order beam scans space in the $z$-direction (inset a).  These beams then pass through AOM2,  where they are again diffracted into beams 1, 2, 3, and 4 (inset b). Beams 1 and 4, respectively, scan space in the y and z-directions.  Beam 3 is a point and beam 2 scans space with a  parabolic path.  An iris is then used to pass only beam 2, whose transverse and axial profiles are shown in inset c. Two sets of relay lenses $(L_1,L_2)$, $(L_3,L_4)$ optically superimpose and collimate the beams so that they share a common origin of angular deflection. A white light source (WLS) provides background illumination for the images of the particles at the CCD2 camera. The 50\% beam splitter (BS) combines laser light with white light. The notch filter NF cuts the laser radiation. The CCD1 camera visualizes the dynamics of the trapped particle in the axial plane (side view)) and CCD2 visualizes it in the transverse plane (front view). To create the non-uniform intensity distribution, the radio frequency of AOM1 is twice the radio frequency of AOM2, and for the uniform distribution, the radio frequencies of both AOMs are equal with a relative phase shift.  The instantaneous power of the scanning beam is $1$ W,  but the particle experiences a $10$ mW averaged power at each point over the parabolic path.  The trapping region is enclosed by a $2 \times 1$ cm$^2$ glass tube to reduce the airflow around the trap site. Particle loading is performed by releasing particles from the top of the glass tube into the trapping region, where they are observed using CCD1 and CCD2. For example, the trapped gold particles are visualized by CCD1 and CCD2 in insets (d) and (e), respectively. (We note that, from the perspective of CCD2, solid gold particles had poor visibility when trapped. Therefore, for illustration purposes, a large gold-coated microsphere was used in this image.)

\begin{figure}
    \centering
     \includegraphics[width=.44\textwidth]{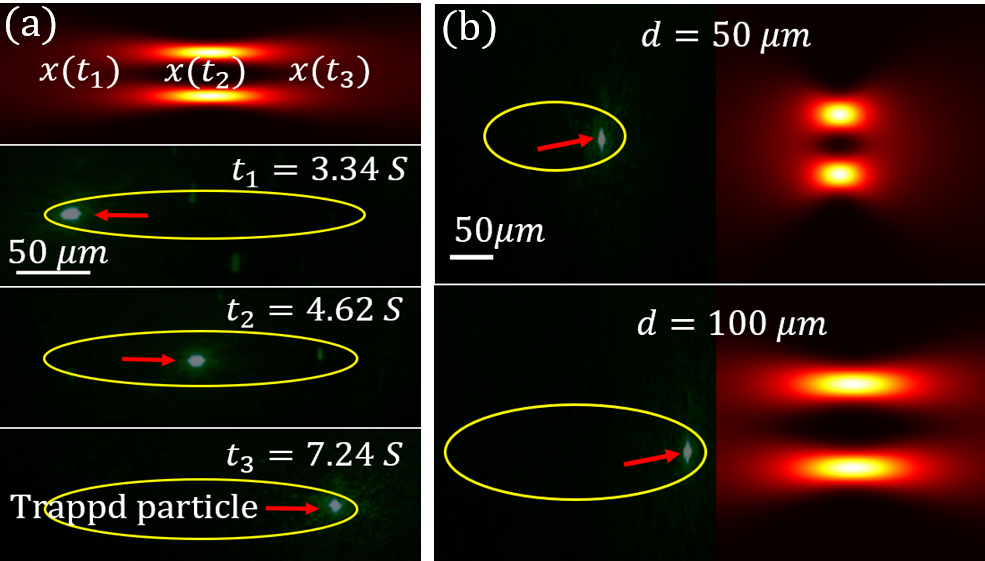}
     \caption{(a) Self-loading operation for uniform intensity distribution at $2$ kHz and $d=50$ $\mu$m (b)  the modulation operation by changing $d$ from $50$ $\mu$m to $100$ $\mu$m.  }
    \label{goldDistributions}
    
\end{figure}

Figure \ref{goldDistributions} (a) and Media 1 show the self-loading operation of particles into the trapping region, resulting in a trapped particle near the positive convergence point as predicted by our theoretical analysis. The laser is scanned at a rate of $2$ kHz in the uniform distribution setting, and the tuning parameter is chosen to be  $d=50$ $\mu$m. The trapping region, where the potential energy well can capture particles, is indicated by the yellow oval. The position of a single particle that is eventually trapped is indicated at different time steps $x(t)$ with red arrows.  When the particle enters the trapping region, it is guided to a single stable trap point as shown by three frames $t_1$, $t_2$, and $t_3$.   This trapping behavior agrees with our theoretical calculations, which predict that in the transverse plane, the particles are confined near the axis of symmetry and in the axial plane are pushed from the left side of the boat trap toward the right side of the boat trap before becoming trapped to the right of the focal plane.

Numerical computation of photophoretic force is computationally intensive due to the interplay of electromagnetic absorption and heat transfer equations \cite{supp}. For instance, calculating temperature distribution for a single particle location required a mesh with 204,747 points and took 9 hours on an 8-core Intel i7-10875H CPU. To mitigate this, we focused on a trap with \(d=5\, \mu m\). Our calculations accurately predict stable points for varying \(d\). For \(d=50\, \mu m\), the positive convergence point is at \(x=106\, \mu m\), aligning well with experimental measurements of \(x=102\, \mu m\). Images are cropped and contrast-enhanced for clarity.

Various trapping parameters can be adiabatically adjusted for real-time 3D control of particle position and velocity. The trap volume depends on the tuning voltage range of the VCO and parameter \(d\), which also affects trapping region dimensions \(H\) and \(L\). In Fig. \ref{goldDistributions}(b) and Media 1, the trap volume increases as \(d\) changes from \(50 \, \mu m\) to \(100 \, \mu m\), altering the trapping point's \(z\)- and \(x\)-coordinates. Experimental observations confirm that increasing \(d\) extends the distance from the trap to the focal point. In addition, the translation operation and trap lifetimes exceeding one hour are discussed in the Supplemental Material \cite{supp}.

\section{Conclusion} 

In this work, we have demonstrated the optical trapping of large metallic particles using a new trapping method that relies on positive photophoretic forces.  The new trap achieves a predictable trap location while preserving high trap probability.  We have further demonstrated how the dynamic nature of this trap may be used to axially scan the trap and increase particle trap count. The results shown establish a trapping approach that resolves the issues afflicting stochastic traps. This advancement allows for more precise particle illumination and manipulation, which is critical for reliable, higher-quality image generation in optical trap display applications and other levitated optomechanical applications.

\begin{acknowledgments}

We acknowledge financial support from the National Science Foundation (Grant No. 2234534). We also thank Spencer Duke for his contributions to the paper.

\end{acknowledgments}

\nocite{*}

%

\end{document}


\preprint{prl/123-QED}
\bibliographystyle{apsrev4-2}

 \author{S. Mirzaei-Ghormish}
\affiliation{Department of Electrical and Computer Engineering, Brigham Young University, Provo, UT}
 
\author{S. Griffith}%
\affiliation{Department of Electrical and Computer Engineering, Brigham Young University, Provo, UT}
  \email{griffithstephen97@yahoo.com}

 \author{D. Smalley}
 \affiliation{Department of Electrical and Computer Engineering, Brigham Young University, Provo, UT}
 \email{smalley.byu.edu}

 \author{Ryan M. Camacho}
 \affiliation{Department of Electrical and Computer Engineering, Brigham Young University, Provo, UT}
 \email{camacho@byu.edu}

\date{\today}

\title{Supplementary Material for ``Optical Trapping of Large Metallic Particles in Air"}

\maketitle

\section{Generation of Uniform Transverse Intensity Distribution of Scanning Laser}

As noted in the main text, the scan speed of the laser must be constant along its parabolic path to generate a uniform quasistatic intensity distribution. Here we give details on how this is accomplished.  

\begin{figure}
    \centering
     \includegraphics[width=\textwidth]{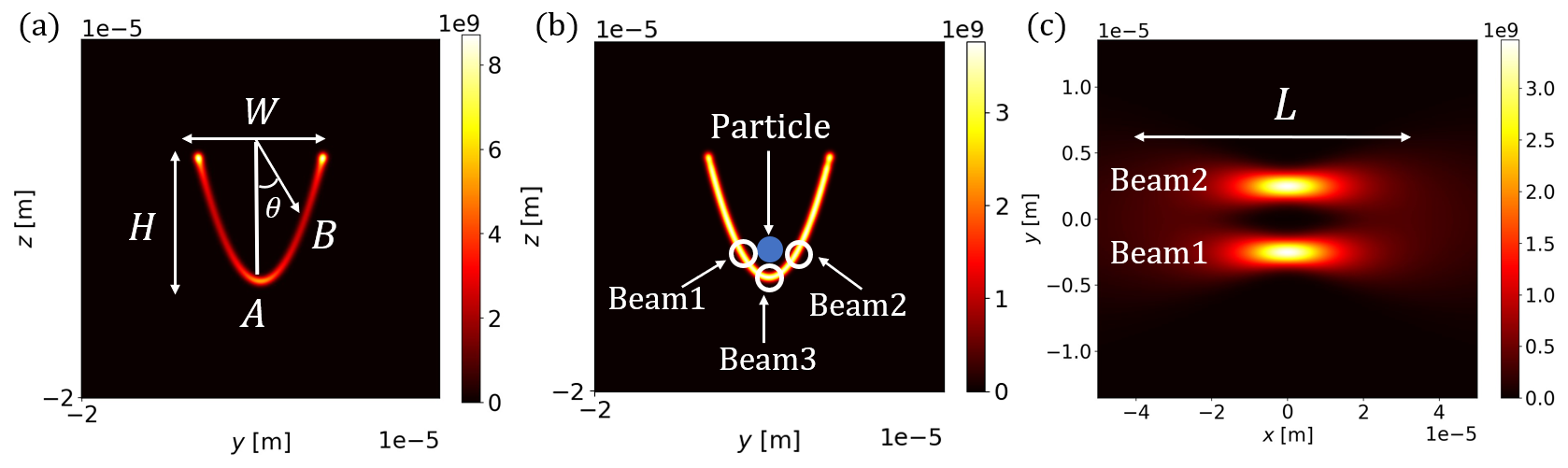}
     \caption{(a) The non-uniform and (b) uniform time-averaged intensity distribution in the transverse plane at $x=0$.   (c) The convergence of beams 1 and 2 along the axial direction. }
    \label{parabola}
    
\end{figure}

The laser beam is linearly polarized in the $z$-direction and propagating in the $+x$ direction. Denoting the scanning frequency is $\omega$, we may write the time-averaged intensity over the time period $T=\frac{2\pi}{\omega}$ as: $I_{ave}(\vec{r})=\frac{1}{T}\int_0^TI(x,y,z;t)dt$, where $ \left(y_f(t),z_f(t)\right)=(d \sin(\omega t), -d \cos(2\omega t))$ and   $\dot{s}(t)=d\omega\sqrt{\cos^2(\omega t)+4\sin^2(2\omega t)}$ is the speed of scanner.
Figure \ref{parabola}(a) shows an example time-averaged non-uniform intensity in the $x=0$ plane, where the simulation parameters are chosen as follows: $\lambda=$ 532 nm, $p_{in}=10$ mW, $NA=0.2$, $\omega=5$ kHz, and $d=5$ $\mu$m. For this intensity distribution, the speed of the scanner is not constant along the parabolic path which would result in a non-uniform distribution, as seen in Fig. \ref{parabola}(a).  As shown experimentally,  the intensity distribution significantly affects the quality of trapping. In the non-uniform case, the lateral walls have a lower average intensity compared to the other parts. Thus, the particles are more likely to leave the trap region from the low-intensity regions before being re-illuminated by the scanning laser beam. 

To overcome this limitation, we create a uniform intensity distribution by using a constant speed for the scanner along the parabolic path. If the scan rate for uniform distribution is the same as the non-uniform case, the speed of the scanner in the uniform distribution is $v=\frac{2s_{tot}}{T}$, in which $s_{tot}=\int_{0}^{T/2}\dot{s}(t)dt$ is the total traveled distance on the parabola path, and $s (t)$ is the position of laser beam at time $t$. To obtain the uniform distribution, a different time scale $t^{'} (t)$ should be used, related to the time scale of the non-uniform distribution $t$ as follows:

  \begin{equation}
      vt^{'}=\int_{0}^{t}\dot{s}(\tau)d\tau.
      \label{equation4}
  \end{equation}

  Then, the uniform time-averaged intensity reads
  \begin{equation}
      I_{ave}(\vec{r})=\frac{1}{t'(T)}\int_0^{t'(T)}I\left(\vec{r};t'(t)\right)dt'.
     \label{equation5}
  \end{equation}

The uniform time-averaged intensity distribution is shown in Fig. \ref{parabola}(b). The distributed intensities located on the opposite sides of the parabola converge on the back ($x<0$) and front sides ($x>0$) in the axial direction as shown in Fig. \ref{parabola}(c). 

\section{{Optical and thermal forces calculations}}
The optical forces are calculated using Maxwell stress tensor (MST) \cite{novotny2012principles}. We first calculate MST ($\bar{\bar{T}}$) by finding the macroscopic electromagnetic field components (incident and scattering) on a closed surface $\partial v$ surrounding the particle.

\begin{equation}
    \bar{\bar{T}}=\varepsilon_0\varepsilon_r\vec{E}\vec{E}+\mu_0\mu_r\vec{H}\vec{H}-\frac{1}{2}\left(\varepsilon_0\varepsilon_r|\vec{E}|^2+\mu_0\mu_r|\vec{H}|^2\right)\bar{\bar{I}}
 \label{equation6}
\end{equation}

where $\vec{E}$ and $\vec{H}$ are the macroscopic electric and magnetic fields, respectively, and $\bar{\bar{I}}$ is the unit tensor. Then, we calculate the time-average force $\langle \vec{F}(\vec{r},t)\rangle$ by integrating MST over $\partial v$. 

\begin{equation}
    \langle \vec{F}_{opt}(\vec{r},t) \rangle =\oint \langle\bar{\bar{T}}(\vec{r},t)\rangle . \hat{n}(\vec{r})ds
    \label{force_time_averaged}
\end{equation}
where $\hat{n}$ is the upward unit vector perpendicular to the surface. We use COMSOL Multiphysics to implement and calculate MST. 

\begin{figure}
    \centering
    \begin{subfigure}{.5\textwidth}
        \includegraphics[width=\textwidth]{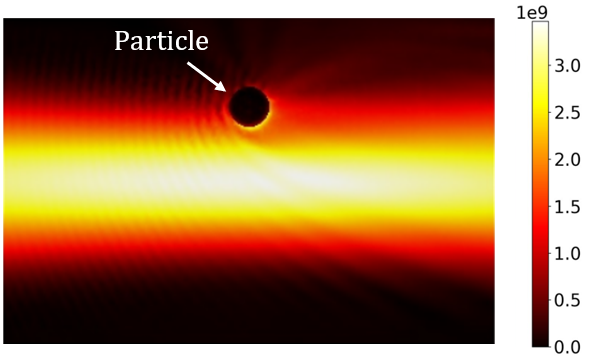}
        \caption{}
        \label{profile}
    \end{subfigure}%
      \centering
    \begin{subfigure}{.5\textwidth}
        \includegraphics[width=\textwidth]{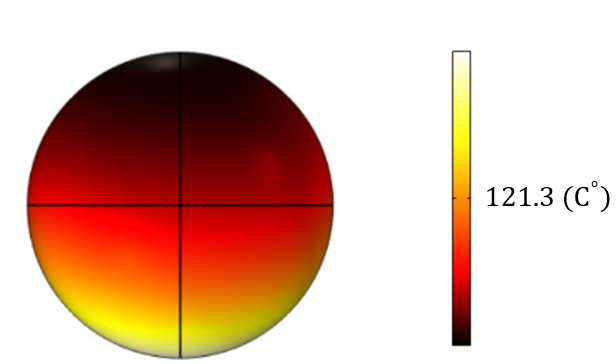}
        \caption{}
        \label{temperature}
    \end{subfigure}
 \caption{(a) The intensity profile of a single laser beam, where a gold microparticle is located on top of the intensity barrier. (b) The temperature distribution on the surface of the particle.}
    \label{intensityDistribution}
\end{figure}

To model the photophoretic force, we note that the mean free path of the surrounding air particles is much smaller than the radius of the trapped particles, allowing us to model photophoretic force in the continuous regime with appropriate boundary conditions.  
We first find the temperature distribution on the particle’s surface using the heat transfer equation:

\begin{equation}
    \vec{\nabla}.\left(\kappa\vec{\nabla}T(\vec{r},t)\right)+Q(\vec{r})=\rho c_p\partial_tT(\vec{r},t)
    \label{equation8}
\end{equation}

where $\kappa$ is thermal conductivity, $T$ is temperature, $\rho$ is mass density, and $c_p$ is specific heat capacity. The volumetric heat source $Q$ is related to the electromagnetic radiation by $Q(\vec{r})=2\pi f\varepsilon_0 \varepsilon_r^{''}I_{ave}(\vec{r})$, where $f$ is laser frequency, and $\varepsilon_r^{''}$ is the imaginary part of the particle’s permittivity. We next apply two boundary conditions to solve the heat transfer equation: absorption of the laser illumination from a given direction, and cooling through thermal emission in all directions.  By combining these boundary conditions, we obtain the Neumann boundary condition 

\begin{equation}
    \kappa\vec{\nabla}T \cdot \hat{n}=I_{ave}\hat{x} \cdot \hat{n}-\sigma\epsilon(T^4-T_{0}^{4}),
    \label{equation9}
\end{equation}
where $T_0=20$ $^{\circ}$C is the ambient temperature, $\sigma$ is the Stefan-Boltzmann constant and $\epsilon$ is emissivity that we set $\epsilon=0.8$. The radius and refractive index of gold particles are $R=500$ nm and $\tilde{N}=0.543+j2.2309$, respectively, and the mean free path of the air at $P=1$ atm  is approximately $69.4$ nm, which results in a Knudsen number of $K_n=0.069$. 
The photophoretic force can be calculated by integrating the inhomogeneous temperature over the particle’s surface 
\cite{zulehner1995representation} 
\begin{equation}
    \vec{F}_{ph}=-\left(\frac{\alpha P}{4T_0}\right)\oint T(\vec{r})\hat{n}ds,
    \label{photophoretic_force}
\end{equation}
where $\alpha$ is thermal accommodation. 
The total force $\vec{F}_{tot}$ is calculated as the sum of the optical force $\vec{F}_{opt}$, the photophoretic force $\vec{F}_{ph}$ and the gravitational force $\vec{F}_{gr}$. To simplify this calculation, we take advantage of the trap symmetry in the transverse $y-z$ plane and the assumption that the quasistatic intensity distribution profile is constant along the parabolic path. The sum of the optical and photophoretic forces $\vec{F}_{opt}$+$\vec{F}_{ph}$  may be found using one force calculation, since the magnitude $\mid \vec{F}_{opt} \mid\ + \mid\vec{F}_{ph}\mid$ is a constant at all points along the parabola, but with a different direction at each point.  We calculate the sum of optical and photophoretic forces $\vec{F}_{A}$ at a point $A= (x,0,z)$ on the $y=0$ plane  and apply a rotation matrix $\bar{\bar{R}}(\theta)$ to find the sum of optical and photophoretic forces $\vec{F}_{B}$  at a point $B=(x,y(\theta),z(\theta))$:

\begin{equation}
    \vec{F}_{B} = \bar{\bar{R}}(\theta)\vec{F}_A=
    \begin{bmatrix}
    1 & 0 & 0\\
    0 & \cos(\theta) & -\sin(\theta)\\
    0 & \sin(\theta) & \cos(\theta)
    \end{bmatrix} 
    \begin{bmatrix}
    F_x\\
    F_y\\
    F_z
    \end{bmatrix}
    , \hspace{.1cm} -\frac{\pi}{2} \leq \theta \leq \frac{\pi}{2}
    \label{equation11}
\end{equation}
where $\vec{F}_{A} = \left[ F_x F_y F_z\right]^T$ and  $(y(\theta),z(\theta))=(d \sin(\theta),-d \cos(2\theta))$. The points $A$, $B$, and angle $\theta$ are specified in Figure \ref{parabola}(a). Figure \ref{intensityDistribution}(a) and \ref{intensityDistribution}(b) show the intensity profile and temperature distribution of the particle’s surface at the same point $A$, which are calculated using COMSOL Multiphysics. In this example, the bottom of the particle is closer to the high-intensity region than the top of the particle, resulting in a higher temperature on the bottom and a photophoretic force in the $+z$ direction.

\begin{figure}
    \centering
    \begin{subfigure}{.5\textwidth}
        \includegraphics[width=\textwidth]{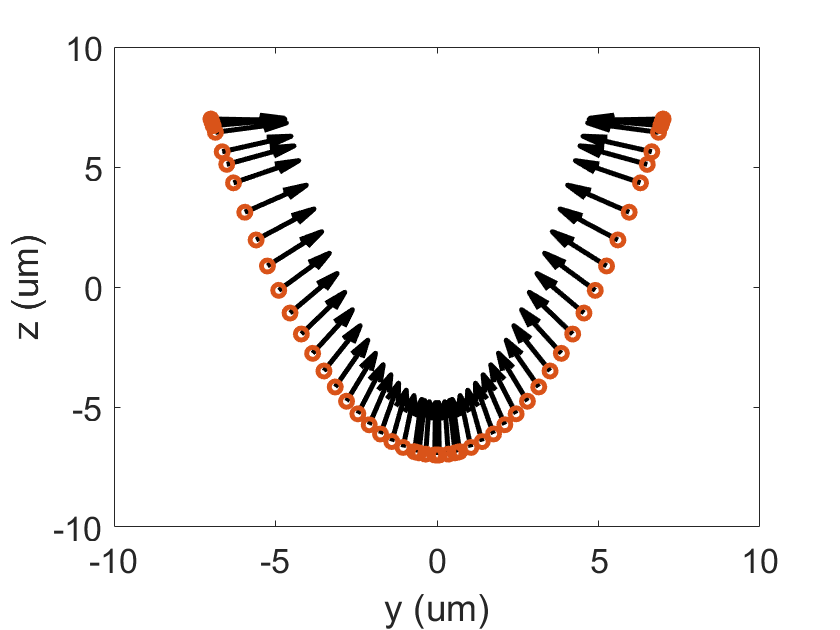}
        \caption{}
        \label{vector force}
    \end{subfigure}%
      \centering
    \begin{subfigure}{.5\textwidth}
        \includegraphics[width=\textwidth]{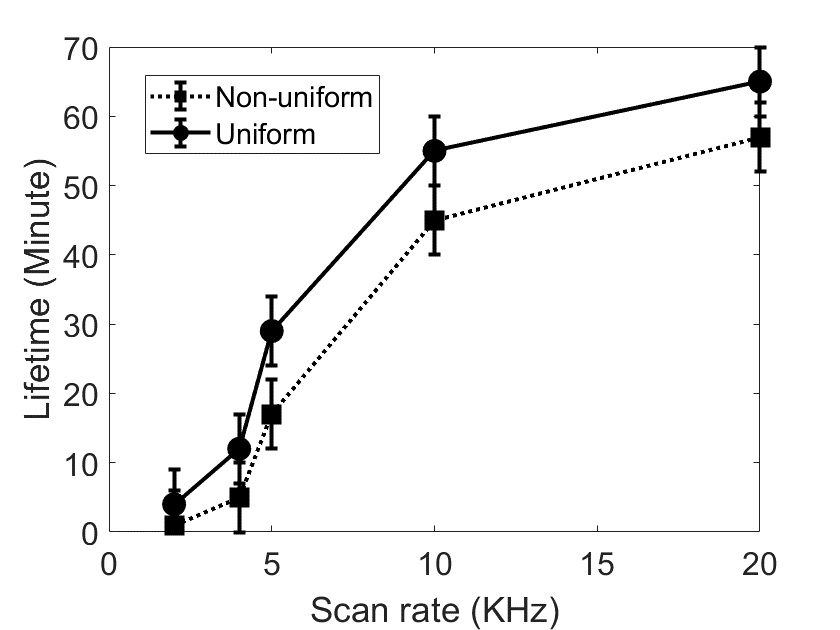}
        \caption{}
        \label{lifetime}
    \end{subfigure}
 \caption{(a) The vector map of the total force in the transverse plane.  (b) The trapping lifetime as a function of scan rate. The solid line shows the lifetime for uniform intensity distribution and the dashed line is the lifetime of the non-uniform intensity distribution.}
    \label{lifetime}
\end{figure}
The vector map of the total force in the transverse plane is shown in Fig. \ref{vector force}. The red circles show the position of the particle on the parabola path and the black arrows indicate the direction and relative magnitudes of the total force.

The mean trapping lifetime as a function of scan rate is shown in Fig. \ref{lifetime}.  Each data point represents the average of 10 experiments, and the bars indicate the standard deviation.  By increasing the scan rate, the lifetime is increased. Moreover, the uniformly distributed intensity beams show better performance than the non-uniform distribution. As shown in the uniform distribution lifetimes beyond 1 hour are achieved at $20$ kHz

\begin{figure}
    \centering
     \includegraphics[width=8cm]{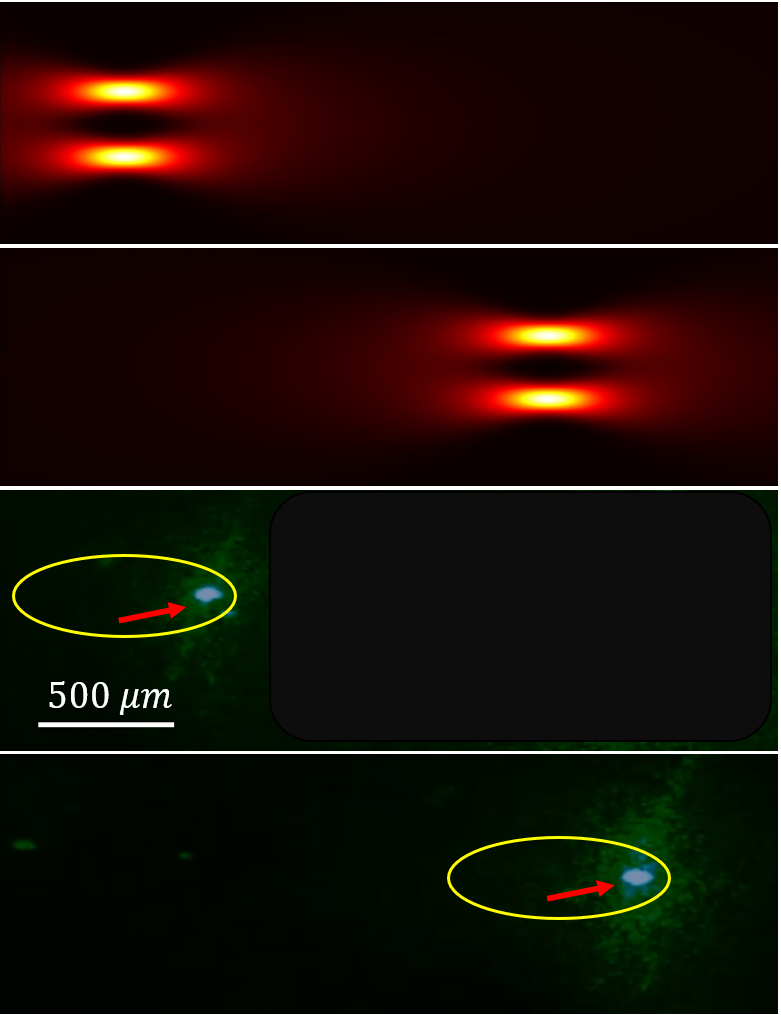}
     \caption{(a) The translation operation for uniform distribution at $10$ kHz.}
    \label{translation}
    
\end{figure}

The trapping locus can easily move in three-dimensional space by shifting the range of the tuning voltage of the VCO.  In Fig. \ref{translation}  the translation of the boat trap along the axial direction is shown for a uniform distribution at 10 kHz. After the particle was stably trapped at the first point, the boat trap site was moved to another location. Consequently, the particle is moved to the second stable point in the axial direction. 

\FloatBarrier
\nocite{*}
\bibliography{Supplementary}